\documentclass{sig-alternate-2013}
\usepackage{url}
\usepackage{proof}
\pdfoutput=1

\newfont{\mycrnotice}{ptmr8t at 7pt}
\newfont{\myconfname}{ptmri8t at 7pt}
\let\crnotice\mycrnotice%
\let\confname\myconfname%

\permission{Permission to make digital or hard copies of all or part
  of this work for personal or classroom use is granted without fee
  provided that copies are not made or distributed for profit or
  commercial advantage and that copies bear this notice and the full
  citation on the first page. Copyrights for components of this work
  owned by others than the author(s) must be honored. Abstracting with
  credit is permitted. To copy otherwise, or republish, to post on
  servers or to redistribute to lists, requires prior specific
  permission and/or a fee. Request permissions from
  permissions@acm.org.}

\conferenceinfo{HILT 2014,}{October 18--21, 2014, Portland, OR, USA.\\
{\mycrnotice{Copyright is held by the owner/author(s). Publication rights licensed to ACM.}}}
\copyrightetc{ACM \the\acmcopyr}
\crdata{978-1-4503-3217-0/14/10\ ...\$15.00.\\
http://dx.doi.org/10.1145/2663171.2663177}

\begin{document}

\title{Resolute: An Assurance Case Language for\\ Architecture Models}
\numberofauthors{2}
\author{
\alignauthor
Andrew Gacek, John Backes,\\ Darren Cofer, Konrad Slind\\
       \affaddr{Rockwell Collins}\\
       \affaddr{Advanced Technology Center}\\
       \email{first.last@rockwellcollins.com}
\alignauthor
Mike Whalen\\
       \affaddr{University of Minnesota}\\
       \affaddr{Computer Science Department}\\
       \email{whalen@cs.umn.edu}
}

\maketitle
\begin{abstract}
Arguments about the safety, security, and correctness of a complex
system are often made in the form of an assurance case. An assurance
case is a structured argument, often represented with a graphical
interface, that presents and supports claims about a system's
behavior. The argument may combine different kinds of evidence to
justify its top level claim. While assurance cases deliver some level
of guarantee of a system's correctness, they lack the rigor that
proofs from formal methods typically provide. Furthermore, changes in
the structure of a model during development may result in
inconsistencies between a design and its assurance case. Our solution
is a framework for automatically generating assurance cases based on
1) a system model specified in an architectural design language, 2) a
set of logical rules expressed in a domain specific language that we
have developed, and 3) the results of other formal analyses that have
been run on the model. We argue that the rigor of these automatically
generated assurance cases exceeds those of traditional assurance case
arguments because of their more formal logical foundation and direct
connection to the architectural model.
\end{abstract}

\category{\\D.2.11}{Software Architectures}{Languages}



\newpage

\section{Introduction}

The design of complex systems such as Unmanned Air Vehicles (UAVs) can
be greatly improved through the use of advanced system and software
architecting tools. In previous work, we have successfully used model
checking to verify software components that have been created using
model-based development (MBD) tools such as Simulink
\cite{miller2010cacm}. An objective of our current research is to
build on this success and extend the reach of model checking and other
formal methods to system design models.

The Architecture Analysis and Design Language (AADL)
\cite{feiler:2012} is targeted for capturing the important design
concepts in real-time distributed embedded systems. The AADL language
can capture both the hardware and software architecture in a
hierarchical format. It provides hardware component models including
processors, buses, memories, and I/O devices, and software component
models including threads, processes, and subprograms. Interfaces for
these components and data flows between components can also be
defined. The language offers a high degree of flexibility in terms of
architecture and component detail. This supports incremental
development where the architecture is refined to increasing levels of
detail and where components can be refined with additional details
over time.

One of our core innovations is to structure the formalizations and
proofs by following the AADL descriptions of the system. In other
work, we did this through the use of formal assume-guarantee contracts
that correspond to the component requirements for each component
\cite{cofer2012nfm}. Our current work on DARPA's High Assurance Cyber
Military Systems (HACMS) program is focused on security properties of
UAVs \cite{hacms}. We have found that in assuring the cyber-security
properties of aircraft designs we need to integrate a variety of
evidence with varying levels of formality. This has been our
motivation to explore assurance case approaches.

In this paper we report on {\it Resolute}, a new assurance case
language and tool which is based on architectural models. In
developing Resolute, we have followed the same approach of embedding
the proof in the architectural model for the vehicle, tightly coupling
terms in the assurance case with evidence derived directly from the
system design artifacts. This ensures that we maintain consistency
between the system design and its associated assurance case(s). Design
changes that might invalidate some aspect of an assurance case can be
immediately flagged by our tool for correction.

\newpage
\section{Assurance Cases}
\label{sec:assurance}

Using Resolute, the goal is to construct an {\em assurance
case} \cite{hawkins2009issc,Kelly:2001sc,kelly98:thesis} about a
system specified in AADL. From~\cite{GSN}, an assurance case is
defined as:
\begin{quote}
A reasoned and compelling argument, supported by a body of evidence,
that a system, service or organization will operate as intended for a
defined application in a defined environment.
\end{quote}
Assurances cases are constructed to show that one or more {\em claims}
about the system are acceptable; usually the claims are defined for an
aspect of the system such as safety ({\em safety cases}) or security
({\em security cases}). For complex systems, these structured
arguments are often large and complicated.  In order to construct, present, 
discuss, and review these arguments, it is necessary that they are clearly 
documented.  Several notations have been proposed to properly 
document assurance cases. The most popular
notation is currently the Goal Structuring Notation
(GSN)~\cite{GSN,Kelly2004:gsn}, which is used in several
assurance-case tool suites.

\begin{figure}
    \includegraphics[trim=50 190 130 100,clip,width=0.5\textwidth]{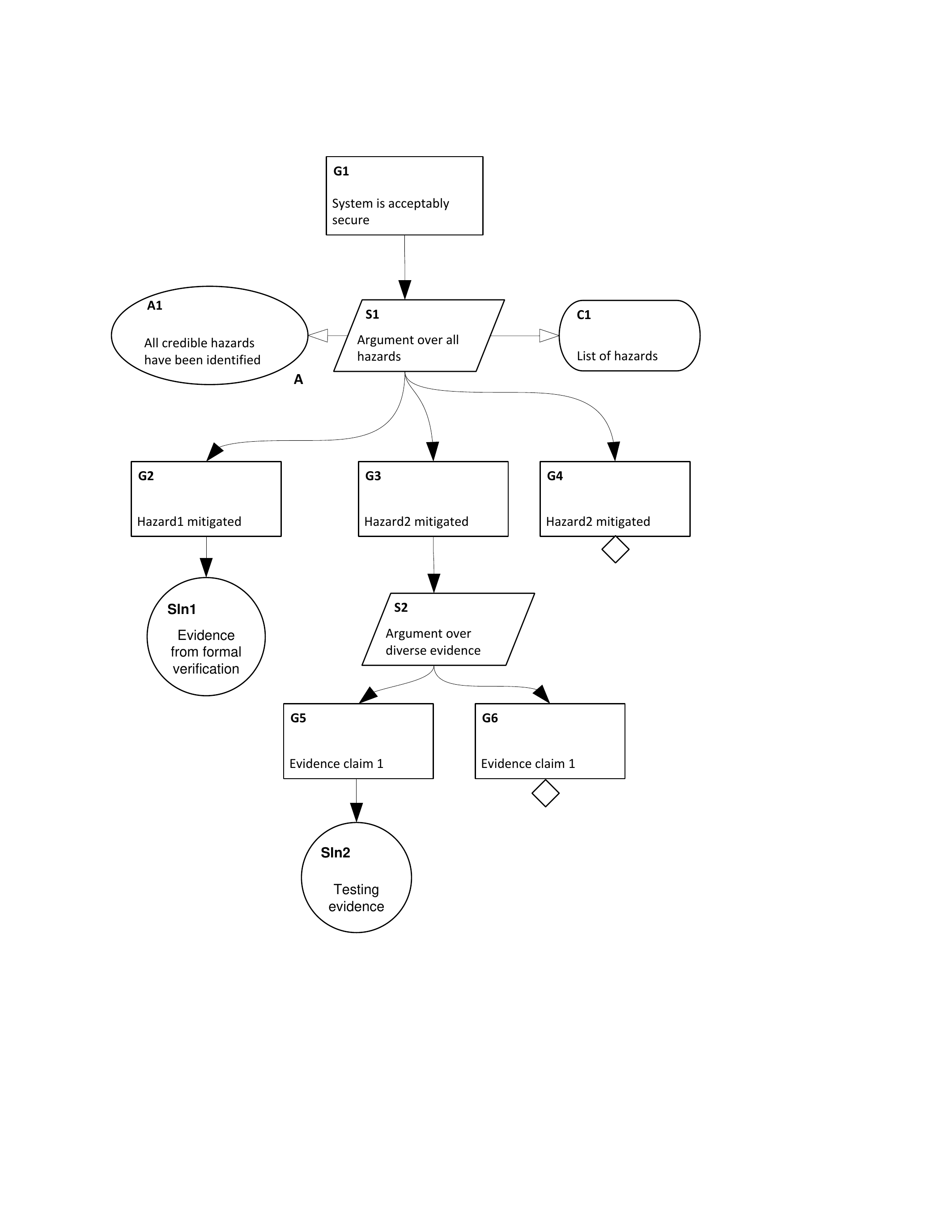}
    \caption{Example of GSN notation}
    \label{fig:gsn}
\end{figure}

An example of the GSN notation is shown in Figure~\ref{fig:gsn}. In
GSN, we have {\em goals} (G1..G6 in Figure~\ref{fig:gsn}) that
represent claims about the system. When a claim is established through
reasoning about subclaims, a {\em strategy} (S1, S2 in
Figure~\ref{fig:gsn}) is used to describe the nature of how the
subclaims establish the claim. Furthermore, strategies often rely on
{\em assumptions} in order for the strategy to be reasonably applied.
In the figure, we make an argument that the system is acceptably
secure by enumerating the hazards that prevent it from being secure
and demonstrating that these hazards are mitigated. An assumption of
this strategy (A1 in Figure~\ref{fig:gsn}) is that we have enumerated
all reasonable security hazards. Also, it is often necessary to
provide the {\em context} in which a strategy or goal occurs: in the
case of the enumeration strategy, the context is the list of
identified hazards. We terminate the argument either in goals that
have no further decomposition, graphically notated by a diamond (shown
in the figure by goals G4 and G6), or in {\em solutions} (Sln1, Sln2
in Figure~\ref{fig:gsn}) that describe evidence for the goal. GSN
arguments form directed acyclic graphs; it is possible to use the same
subclaim or evidence as part of the justification of a larger claim,
but it is not well-formed to have a cyclic chain of reasoning within a
GSN graph.

To be compelling, the argument must provide sufficient assurance in
the claims made about the system. Constructing such arguments is quite
difficult, even given appropriate notations. First, proper claims must
be identified to define the objectives of the assurance case. Then
appropriate argumentation must be constructed: such argumentation must
often take into account the environment in which the system is used,
the artifacts that are constructed during system design,
implementation, and test, and the processes followed during the
development and implementation cycle. There is a rich body of
literature that describes proper processes and patterns for
constructing assurance cases, {\it e.g.}~\cite{hawkins2009issc,
Kelly:2001sc, kelly98:thesis, Kelly97:patterns, Sun11:pattern,
Denney13:pattern, Hawkins11:pattern, Greenwell06:fallacies,
Graydon07:dev}.

\section{AADL}

\begin{figure*}[t]
\includegraphics[scale=0.7]{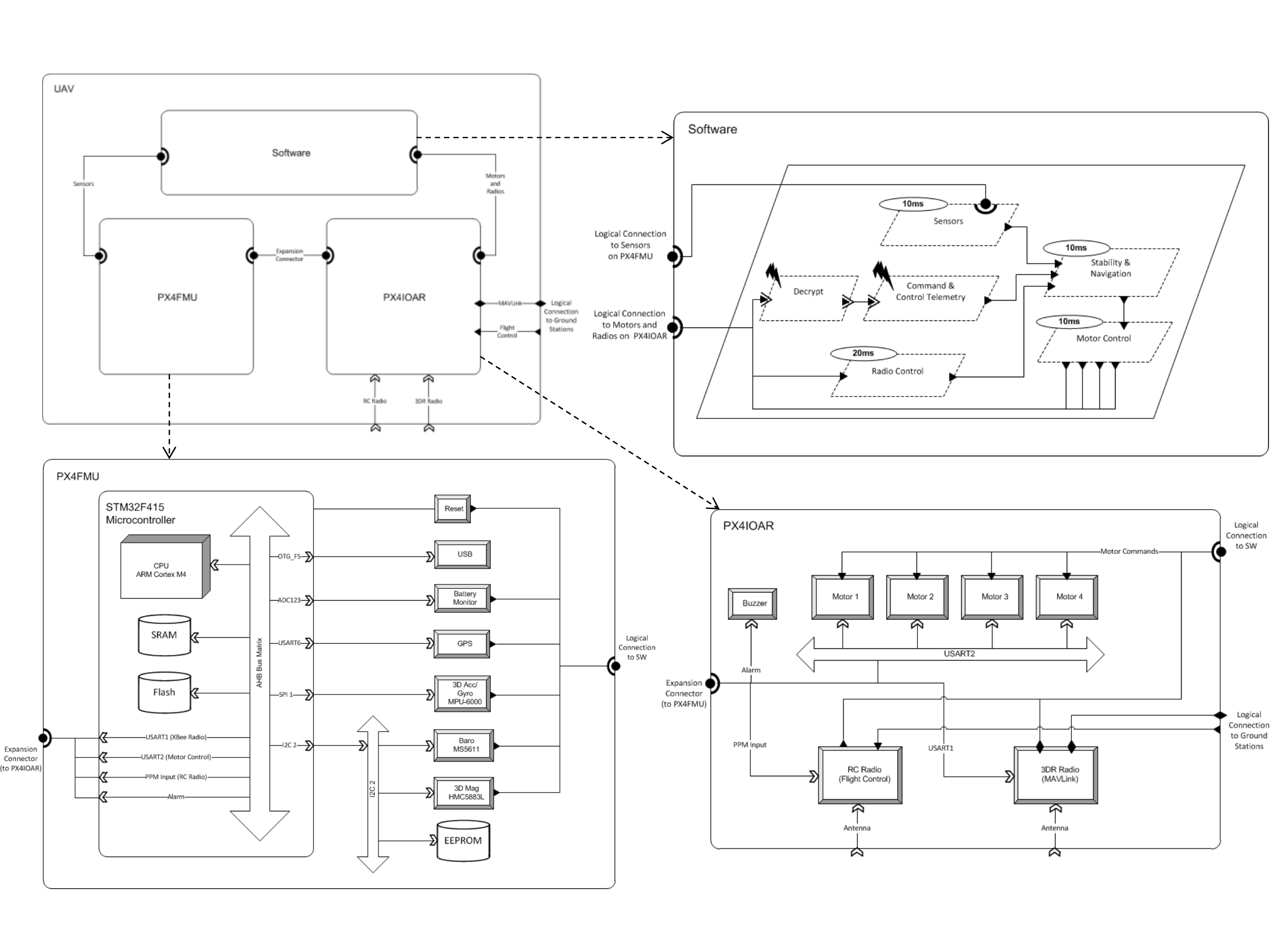}
\caption{Simplified AADL model of HACMS quadcopter}
\label{fig:aadl}
\end{figure*}

Our domain of interest is distributed real-time embedded systems
(including both hardware and software), such as comprise the critical
functionality in commercial and military aircraft. Many aerospace
companies have adopted MBD processes for production of software
components, and we have successfully applied formal analysis at this
level to verify component requirements. However, the system design
process is often less rigorous: system-level descriptions of the
interactions of distributed components, resource allocation decisions,
and communication mechanisms between components are typically ad hoc
and not based on analyzable models. Application of formal analysis
methods at the system level requires 1) an abstraction that defines
how components will be represented in the system model, and 2)
selection of an appropriate formal modeling language.

In this approach, the architectural model includes interface,
interconnections, and specifications for components but not their
implementations. It describes the interactions between components and
their arrangement in the system, but the components themselves are
black boxes. The component implementations are described separately by
the existing MBD environment and artifacts (or by traditional
programming languages, where applicable). They are represented in the
system model by the subset of their specifications that is necessary
to describe their system-level interactions. This distinction between
system architecture and component implementation is important to
ensure the scalability of the analyses that we wish to perform.

We have selected AADL as a system modeling language for our work. AADL
is described in SAE standard AS5506B and has a sufficiently precise
definition to support formalization of its semantics
\cite{feiler:2012}. It provides syntax for describing both hardware
and software aspects of the system so that requirements related to
resource allocation, scheduling, and communication between distributed
elements can be addressed. Textual and graphical versions of the
language are available with tool support for each \cite{osate}. An
important feature of the language is its extensibility via the annex
mechanism. Language developers can embed new syntax into an AADL model
to provide new features or to support additional analyses. Annex
expressions have full access to the rest of the model providing the
ability to refer to components and properties described in the base
AADL language. We have used this annex mechanism to add behavioral
contracts and assurance case rules to our system architecture models.

Figure~\ref{fig:aadl} shows an example AADL system model (using the
graphical syntax) that we have developed for the HACMS program. The
model describes a UAV that includes a simplified version of the system
software along with the processing hardware (PX4FMU) and an I/O board
for motors and radios (PX4IOAR). The actual flight software model is
over 7,000 lines of AADL and includes 35 computing threads, with a C
implementation of over 50K lines of code.

In the HACMS project we are concerned primarily with cyber-security of
UAVs. We are developing analysis tools based on the AADL language that
allow us to verify important security and information flow properties
for this kind of system. We are also developing synthesis tools that
allow us to generate glue code and system configuration data for the
system. This data, along with the component implementations, provide
everything needed to build the final flight binaries from the AADL
model. To show that our results scale to real military systems, we are
also transitioning these technologies for demonstration on Boeing's
Unmanned Little Bird helicopter \cite{ulb}.

\section{Resolute}

Resolute is a language and tool for constructing assurance cases based
on AADL models. Users formulate claims and rules for justifying those
claims, which Resolute uses to construct assurance cases. Both the
claims and rules are parameterized by variable inputs which are
instantiated using elements from the models. This creates a dependence
of the assurance case on the AADL model and means that changes to the
AADL model can result in changes to the assurance case. This also
means that a small set of rules can result in a large assurance case
since each rule may be applied multiple times to different parts of
the architecture model.

Resolute is designed primarily to show the structure of an argument;
in the GSN notation, it would define how a claim is provable from
subclaims. In some ways, Resolute is richer than GSN; as shown below,
it supports parametric goals and arbitrary Boolean relations between
claims and subclaims rather than simple conjunctions or $m$-of-$n$
relations. On the other hand, it does not currently provide specific
placeholders for {\em context} and {\em assumption} information. In
GSN, this information is explicit, but informal: it is documented to
aid the readers and writers of the assurance case. Resolute currently
does not have a placeholder for this information, but this could
easily be remedied by adding string properties to document contextual
aspects of claim/subclaim relationships. To define context and
assumption ideas more formally is more challenging and something that
we are considering for future work.

\subsection{Claims and Rules}

In Resolute, each claim corresponds to a first-order predicate. For
example, a user might represent a claim such as ``The memory of
process {\tt p} is protected from alteration by other processes''
using the predicate {\tt memory\_protected(p : process)}. The user
specifies rules for {\tt memory\_protected} which provide
possible ways to justify the underlying claim. Logically, these rules
correspond to global assumptions which have the form of an implication
with the predicate of interest as the conclusion. For example, an
operating system such as NICTA's secure microkernel seL4---which we
are using in HACMS---might enforce memory protection on its own
\cite{klein2009sigops}:
\begin{verbatim}
  memory_protected(p : process) <=
    (property_lookup(p, OS) = "seL4")
\end{verbatim}
Here we query the architectural model to determine the operating
system for the given process. Another way to satisfy memory protection
may be to examine all the other processes which share the same
underlying memory component. Note that in AADL a ``process'' represents
a logical memory space while a ``memory'' represents a physical memory
space.
\begin{verbatim}
  memory_protected(p : process) <=
    forall (mem : memory). bound(p, mem) =>
      forall (q : process). bound(q, mem) =>
        memory_safe_process(q)
\end{verbatim}
In the above rule, we are querying the architectural model via the
universal quantification over memory and process components. Note that
quantification is always finite since we only quantify over
architectural components and other finite sets. The built-in {\tt
  bound} predicate determines how software maps to hardware. In
addition, we call another user defined predicate {\tt
  memory\_safe\_process} to determine if a process is memory safe. In
the resulting assurance case, the claim that a process {\tt p} is
memory protected will be supported by subclaims that all processes in
its memory space are memory safe. Thus there will be one supporting
subclaim for each process in the memory space.

The above rules for memory protection illustrate a couple of ways to
justify the desired claim, but they do not constitute a complete
description of memory protection nor a complete listing of sufficient
evidence. This is a critical point in Resolute: rules are sufficient,
but not complete. The negation of a claim can never be used in an
argument ({\it i.e.}, in logic programming parlance, we do not make a
closed world assumption). This is a manifestation of the traditional
phrase ``absence of evidence is not evidence of absence.'' Instead, if
the user truly wants to use a claim in a negative context, that
notion must be formalized as a separate positive claim with its own
rules for what constitutes sufficient evidence. For example, one may
be interested in a claim such as {\tt memory\_violated} which has
rules which succeed only when a concrete memory violation is detected.

\subsection{Computations}

Separate from claims, Resolute has a notion of {\it computations}
which are complete and can thus be used in both positive and negative
contexts. Usually these computations are based on querying the model.
For example, the {\tt bound} predicate above is a built-in computation
which returns a Boolean value and is used in a negative context in the
rule for {\tt memory\_protected}. Users may also introduce their own
functions which are defined via a single equation such as
\begin{verbatim}
  message_delay(p : process) =
    sum({thread_message_delay(t)
           for (t : thread) if bound(t, p)})
\end{verbatim}
Here {\tt sum} is a built-in function and {\tt thread\_message\_delay}
is another user-defined function.

Computations may contribute to an assurance case, but they do not
appear in it independently since they do not make any explicit claim.
Instead, a user may wrap claims around computations as needed, for
instance a claim such as ``message delay time for {\tt p} is
within acceptable bounds'' using the {\tt message\_delay} function.

Since claims cannot be used negatively while computations can, claims
may not appear within computations. This creates two separate levels
in Resolute: the logical level on top and the computation level
beneath it. The logical level determines the claims, rules, and
evidence used in the assurance case argument, while the computation
level helps determine which claims are relevant in a particular
context and may directly satisfy some claims by performing
computations over the model.

External analyses are incorporated in Resolute as computations. An
external analysis is run each time the corresponding computation is
invoked. This is useful for deploying existing tools for analyzing
properties such as schedulability or resource allocation.

\section{Tool Environment}
\label{sec:tools}

\begin{figure*}[t]
\begin{verbatim}
only_receive_decrypt(x : component) <=
  ** "The component " x " only receives messages that pass Decrypt" **
  forall (c : connection).
    (parent(destination(c)) = x) =>
      is_sensor_data(c) or only_receive_decrypt_connection(c)

only_receive_decrypt_connection(c : connection) <=
  ** "The connection " c " only carries messages that pass Decrypt" **
  let src : component = parent(source(c));
  unalterable_connection(c) and (is_decrypt(src) or only_receive_decrypt(src))
\end{verbatim}
\caption{Example Resolute rules}
\label{fig:ex-rules}
\end{figure*}

\begin{figure*}[t]
\begin{verbatim}
bound(logical : component, physical : component) : bool =
  memory_bound(logical, physical) or
  connection_bound(logical, physical) or
  processor_bound(logical, physical)

memory_bound(logical : component, physical : component) : bool =
  has_property(logical, Deployment_Properties::Actual_Memory_Binding) and
  member(physical, property(logical, Deployment_Properties::Actual_Memory_Binding))

connection_bound(logical : component, physical : component) : bool =
  has_property(logical, Deployment_Properties::Actual_Connection_Binding) and
  member(physical, property(logical, Deployment_Properties::Actual_Connection_Binding))

processor_bound(logical : component, physical : component) : bool =
  has_property(logical, Deployment_Properties::Actual_Processor_Binding) and
  member(physical, property(logical, Deployment_Properties::Actual_Processor_Binding))
\end{verbatim}
\caption{Definition of {\tt bound} in the Resolute standard library}
\label{fig:bound}
\end{figure*}

We have implemented Resolute as an AADL annex using the Open Source
AADL Tool Environment (OSATE) \cite{osate} plug-in for the Eclipse
IDE. Resolute itself is Open Source under a BSD License and available
online \cite{resolute}. Using OSATE, users are able to interact with
Resolute in the same environment in which they develop their AADL
models. In addition, the resulting framework provides on-the-fly
syntactic and semantic validation. For example, references to AADL
model elements in the Resolute annex are linked to the actual AADL
objects in the same project so that undefined references and type
errors are detected instantly.

The syntax of Resolute is inspired by logic programming. Each rule
defines the meaning and evidence for a claim. The meaning of a claim
is given by a text string in the rule which is parameterized by the
arguments of the claim. The body of the rule consists of an expression
which describes sufficient evidence to satisfy that claim. Claims
may be parameterized by AADL types (e.g., threads, systems, memories,
connections, etc.), integers, strings, Booleans, or sets.

Figure~\ref{fig:ex-rules} shows an example of two Resolute rules. The
meaning of the claim is given by the associated text, for example {\tt
  only\_receive\_decrypt(x)} means: ``The component {\tt x} only
receives commands that pass Decrypt.'' An instantiated version of this
string is what will appear in the corresponding assurance case. The
built-in functions like {\tt destination} and {\tt source} return the
feature to which a connection is attached, and the built-in {\tt
  parent} then gives the component which holds that feature. These
rules also make use of other user-defined claims such as {\tt
  is\_sensor\_data} and {\tt unalterable\_connection} which talk about
the content and integrity of connections. Note that the two claims
shown in the figure are mutually recursive. Together, these claims
walk over a model cataloging the data-flow and constructing a
corresponding assurance case.

Many claims, rules, and functions will appear within a Resolute annex
library which is typically a top-level file in an AADL project. These
libraries define the rules for all claims in Resolute, but do not make
any assertions about what arguments the claims should hold on. In
addition, Resolute comes with a standard library of predefined
functions for common operations. For instance, the {\tt bound}
predicate for determining if a logical component is bound to a specific
physical component is part of the standard library and defined as in
Figure~\ref{fig:bound}.

An assurance case is initiated in Resolute by adding a {\it prove}
statement to the Resolute annex for an AADL component. A prove
statement consists of a claim applied to some concrete arguments. An
example prove statement is shown in Figure~\ref{fig:prove} where the
claim {\tt only\_receive\_ground\_station} is associated with the
motor controller thread. When a Resolute analysis is run on an AADL
system instance, an assurance case is generated for every prove
statement that appears in any component within that instance.

\begin{figure}
\begin{verbatim}
process implementation Main_Loop.Impl
  subcomponents
    MC: thread Motor_Control

  ...

  annex resolute {**
    prove only_receive_ground_station(MC)
  **}
end Main_Loop.Impl;
\end{verbatim}
\caption{Prove statements for Resolute claims}
\label{fig:prove}
\end{figure}

\begin{figure*}[t]
\includegraphics[scale=0.75]{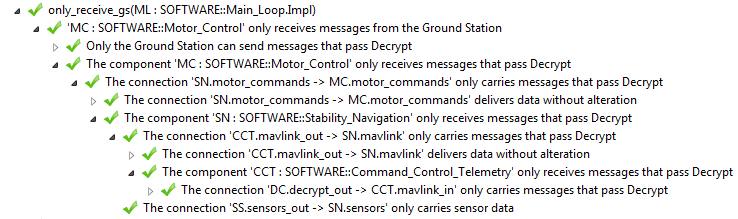}
\caption{Example of a successful assurance case from Resolute}
\label{fig:ex-success}
\vspace{0.2in} 
\end{figure*}

\begin{figure*}[t]
\includegraphics[scale=0.75]{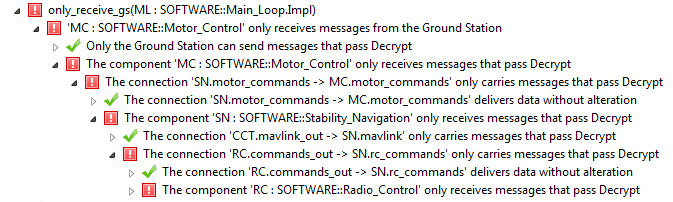}
\caption{Example of a failed assurance cases from Resolute}
\label{fig:ex-failure}
\end{figure*}

Figure~\ref{fig:ex-success} shows a portion of a successful assurance
case generated by Resolute on our simplified UAV model. Each claim is
shown on a single line. Supporting claims are shown indented one level
beneath the claim they support. A check next to a claim indicates that
it is proven. Figure~\ref{fig:ex-failure} shows a portion of a failed
assurance case. An exclamation point indicates that a claim has
failed. In this case, the AADL model includes a safety controller
which is allowed to bypass the Decrypt component and directly send
messages to the UAV. This bypass is detected Resolute. In fact, the
only difference between Figures~\ref{fig:ex-success} and
\ref{fig:ex-failure} is the AADL model. The claims and rules are
identical in both.

The assurance case shown in Figure~\ref{fig:ex-success} is constructed
over our simplified UAV model. We ported this assurance case to the
true UAV model once the latter was available. Although the true UAV
model contained seven times as many software components as the
simplified model, very few of the Resolute rules needed to be changed.
The most significant change was that the true UAV model has data-flow
cycles, and therefore the simple recursive rules used in
Figure~\ref{fig:ex-rules} are insufficient. Instead, we created more
sophisticated rules which recursively computed the set of components
which were reachable prior to passing through the Decrypt component,
and then we justified the claim that the given set was complete and
did not have access to the motor control component.

Assurance cases as shown in Figures~\ref{fig:ex-success} and
\ref{fig:ex-failure} are interactive in the Resolute user interface.
The user can navigate through the assurance case and select a claim to
navigate to locations in the model relevant to the claim. For example,
the user can navigate to any of the AADL components referenced as
input parameters to the claim or can navigate to the rule that defines
the claim. This makes it much easier to figure out why an assurance
case is failing or why a particular part of the assurance case has a
given structure.

An assurance case generated by Resolute is also a stand-alone object.
After construction, it no longer depends on Resolute or even the AADL
model, though it of course still refers to elements of the model. This
means the assurance case can be used as an independent certification
artifact. In addition, Resolute allows assurance cases to be exported
to other formats and assurance case tools such as CertWare
\cite{certware}.

\section{Formal Logic}

\begin{figure*}[t]
\begin{align*}
\infer[\land\mathcal{R}]
      {\Gamma \vdash G_1 \land G_2}
      {\Gamma \vdash G_1 & \Gamma \vdash G_2}
&&
\infer[\lor\mathcal{R}_{i=1,2}]
      {\Gamma \vdash G_1 \lor G_2}
      {\Gamma \vdash G_i}
&&
\infer[\Rightarrow\mathcal{R}]
      {\Gamma \vdash G_1 \Rightarrow G_2}
      {\Gamma, G_1 \vdash G_2}
\end{align*}
\begin{align*}
\infer[\forall\mathcal{R}]
      {\Gamma \vdash \forall x\! :\! \alpha.~ G(x)}
      {\Gamma \vdash G(t_1) & \cdots & \Gamma \vdash G(t_n)}
&&
\infer[\exists\mathcal{R}_{i=1\ldots n}]
      {\Gamma \vdash \exists x\! :\! \alpha.~ G(x)}
      {\Gamma \vdash G(t_i)}
\end{align*}
\begin{center}
where $\alpha = \{t_1, \ldots, t_n\}$
\end{center}
\begin{align*}
\infer[\mbox{backchain}]
      {\Gamma \vdash A(\bar{t})}
      {\Gamma \vdash G(\bar{t})}
\end{align*}
\begin{center}
where $\forall \bar{x}. A(\bar{x}) \Leftarrow G(\bar{x}) \in \Gamma$.
\end{center}
\begin{align*}
\infer[\mbox{eval-}\mathcal{L}]
      {\Gamma, \langle e \rangle \vdash G}
      {e \Downarrow \mbox{false}}
&&
\infer[\mbox{eval-}\mathcal{R}]
      {\Gamma \vdash \langle e \rangle}
      {e \Downarrow \mbox{true}}
\end{align*}
\caption{Resolute logic rules}
\label{fig:logic-rules}
\end{figure*}

The Resolute language consists of both a logic and a computational
sublanguage. The logic of Resolute is an intuitionistic logic similar
to pure Prolog, but augmented with explicit quantification. The logic
is parameterized by the computational sublanguage, and requires only
that the sublanguage is deterministic. This allows the computational
sublanguage to be customized to any domain (e.g. AADL in our context)
and to be expanded and refined, without worrying about the logical
consequences. In fact, we do not even require termination for the
computational sublanguage, though in practice a non-terminating
computation will lead to non-terminating proof search.

\subsection{Syntax}

Let the type of formulas be $o$. We assume the usual logical
constants of $\land, \lor, \Rightarrow : o \to o \to o$ and $\forall,
\exists : (\alpha \to o) \to o$ for every type $\alpha$ not containing
$o$. Let the type of Booleans be $bool$ with constants $true$ and
$false$. We use the constant $\langle \cdot \rangle : bool \to o$ to
inject Booleans into formulas. We assume a notion of evaluation $e
\Downarrow v$ read ``$e$ evaluates to $v$''. We assume evaluation is
deterministic. The full set of types and terms is left unspecified,
but would typically be determined by the computational sublanguage.

\subsection{Sequent Rules}

We define a judgment $\Gamma \vdash G$ where $\Gamma$ is a set of
formulas called assumptions and $G$ is a formula called the goal. This
judgment holds when the goal $G$ is a consequence of assumptions
$\Gamma$ in the Resolute logic. The rules for this judgment are
presented in Figure~\ref{fig:logic-rules}.

The rules for $\land$, $\lor$, and $\Rightarrow$ are standard. The
quantification rules apply only to types with finitely many
inhabitants since the rules work via explicit enumeration. This
simplistic treatment of quantification means that proof search only
needs to consider ground terms. Moreover, finiteness is appropriate
for our domain where we want to quantify over types such as all
threads in a model or all processes within a particular system. The
rule for backchain allows the assumptions in $\Gamma$ to be used in
constructing a proof of an atomic goal. Note that in the backchain
rule, $A$ stands for an atomic formula, {\it i.e.}, a predicate
applied to arguments. Finally, the rules for evaluation allow a proof
to be finished by finding an assumption which evaluates to $false$ or
a conclusion that evaluates to $true$.

User specified claims in Resolute are treated as predicates in the
logic, and the rules for claims are treated as initial assumptions.
Each prove statement in the AADL model is translated to a goal $G$
while all the Resolute rules are translated into the initial context
$\Gamma$. Then proof search is performed on $\Gamma \vdash G$. If a
proof is found, that proof is transformed into an assurance case by
replacing each intermediate sequent of the form $\Gamma \vdash
A(\bar{t})$ by the instantiated version of the claim text for the
claim $A(\bar{x})$. Thus, for us, an assurance case is a proof in the
Resolute logic, and browsing the assurance case means traversing the
proof tree.

\subsection{Customizing the Resolute Logic for AADL}

The Resolute logic we use in our implementation is a customization of
the general Resolute logic. In particular, we allow quantification
over all AADL model types (threads, process, etc) and over all user
computed sets. Our computational sublanguage is based on the
Requirements Enforcement Analysis Language \cite{gilles2010iceccs}.
Our sublanguage includes all the AADL model types and more traditional
types of integers, reals, strings, ranges, and sets. There are
pre-defined functions for common operations (e.g. sum, union, member)
or queries against the AADL model properties and components. Users may
also define their own functions even using recursion, and thus our
computational sublanguage is Turing complete. Moreover, our
sublanguage allows calls out to external tools for richer analyses
such as scheduling or model checking.

Users may specify any rules or meanings for claims, and thus Resolute
can make no judgment about how valid the resulting argument is.
Resolute only ensures that the constructed assurance cases are valid
with respect to the user specified claims and rules. Ultimately, the
acceptability of an assurance case generated by Resolute must depend
on traditional assurance case techniques such as expert review.
Resolute provides a way of keeping an assurance case synchronized with
an architecture model, but the quality of that assurance case is still
dependent on the user.

\section{Related Work}

As discussed in Section~\ref{sec:assurance}, assurance cases have a
large and well-developed literature. Patterns for assurance case
argumentation have been considered
in~\cite{Kelly97:patterns,Sun11:pattern,Denney13:pattern,Hawkins11:pattern},
and common fallacies in assurance cases are considered
in~\cite{Greenwell06:fallacies}. An approach to apply and evolve
assurance cases as part of system design is found
in~\cite{Graydon07:dev}, which is similar to the process we have used
in applying the Resolute tools. A comparison of assurance cases to
prescriptive standards such as DO178B/C is provided by
\cite{Hawkins13ss}. Recent work on {\em confidence cases} as a means
of assessing assurance case arguments is found
in~\cite{goodenough12:confidence}.

Several commercial and research tools support the development of
assurance cases. ASCE~\cite{asce} from Adelard is currently the most
widely used commercial tool for constructing assurance cases. ASCE
supports integration with commercial requirements management tools
such as DOORS, constructing confidence cases with assurance cases, and
integration with a variety of tools through its plug-in architecture.
Other assurance case tools include AdvoCATE~\cite{denny2012safecomp}
from NASA Ames, CertWare~\cite{certware} from NASA Langley,
D-Case~\cite{matsuno2010hase}, and NOR-STA~\cite{cyra2007depcos}.
These tools provide structured editing, visualization, metrics, and
reasoning tools for safety arguments but are not tied into a system
architectural model.

In~\cite{bishop13:tap}, a safety case in ASCE involving a combination
of mechanized proof, testing, and hand-proofs is used to argue that
the maximum error introduced in the computation of a monotonic
function is within some tolerance of the actual value of the function.
This is similar to how we use Resolute in the example in
Section~\ref{sec:tools}; we assemble disparate evidence from different
verification techniques towards an argument. Unlike our work, the ASCE
safety case is not directly integrated into the software/system
architecture. Similar work in~\cite{denny2013:evidence} describes
patterns for using proofs within a safety case and automation for
generating portions of the proof as a part of the safety case.

The Evidential Tool Bus (ETB)~\cite{cruanes13:etb} is very similar in
syntax and semantics to Resolute. It is supported by a Datalog-style
logic and is designed to combine evidence from a variety of sources.
However, the focus of the ETB is on distribution and on {\em
  provenance}---that is, to log the sequence of tool invocations that were
performed to solve the query. It uses timestamps to determine which
analyses are out of date with respect to the current development
artifacts and to only re-run those analyses that are not synchronized
with the current development artifacts. In addition, it is designed to
perform distributed execution of analyses. Analysis tool plug-ins are
used to execute the analysis tools within ETB. ETB is designed to be
tool and model agnostic, and is therefore not integrated with a model
of the system architecture.

The work in \cite{basir2010safecomp} ties together an assurance case
with a model-based notation (Simulink) for the purpose of
demonstrating that the Simulink-generated code meets its requirements.
This work has many similarities to ours, in that the assurance case is
closely tied to the hierarchical structure of the model. It is more
rigorous (in that the assurance case is derived from a formal proof)
but also much more narrow, corresponding to a component in the system
assurance cases that we create. The two approaches could perhaps be
integrated to provide more rigorous safety cases for a wider class of
software developed in a model-based environment.

\section{Future Work}
We have generated a number of assurance cases with Resolute for the
design of a UAV in the HACMS project. Specifically, we have generated
assurance cases that reason about the flow of information through the
vehicle and the availability of resources under different operating
modes. The tool has been useful for modeling requirements of the
architecture at early phases of the design, and verifying that they
still hold in later phases. In future work we plan to make the
assurance cases generated by Resolute exportable to more assurance
case tools. In order to support this, we may extend Resolute with a
more complete set of standard assurance case constructs. For example,
we may introduce strategies as first-class constructs by augmenting
Resolute rules with explicit textual descriptions that would then
appear in the assurance case. We also plan to improve the grammar to
support more features of AADL.

\section{Acknowledgments}
The work presented here was sponsored by DARPA as part of the HACMS
program under contract FA8750-12-9-0179.

\end{document}